%Paper: hep-ph/9205245
%From: "Jiang Liu" <JIANGLIU@penndrls.upenn.edu>
%Date: Fri, 29 May 92 16:37:44 EST

\input phyzzx
\PHYSREV
\hbox to 6.5truein{\hfill UPR-0508T}
\hbox to 6.5truein{\hfill UM-TH-92-13}
\hbox to 6.5truein{\hfill May 1992}
\title{
QCD CORRECTIONS TO THE CHARGED-HIGGS-BOSON DECAY OF A HEAVY TOP QUARK}
\author{Jiang Liu}
\address{Department of Physics, University of Pennsylvania, Philadelphia,
         PA 19104}
\author{York-Peng Yao}
\address{Randall Laboratory of Physics, University of Michigan, Ann
Arbor,  MI 48109}
$$ $$
\abstract
It is shown that up to an over all scale
the lowest-order QCD corrections to  $t\to H^+b$
and to  $t\to W^+b$  are the same in the heavy top limit.
Asymptotically, they are given by
$-{4\alpha_s\over 3\pi}[{\pi^2\over 3}-{5\over 4}]$, resulting in
 a reduction in the decay rate by about $9\%$, rather than
$6\%$ reported previously in the literature.
This  is  verified explicitly
by an analytic calculation.  The application of the equivalence
theorem to this  process is also discussed.

PACS\# 12.38.Bx
\endpage

\chapter{Introduction}

If the top quark $t$ decays according
to the standard model, the CDF experiment has set a lower
limit\Ref\CDF{F. Abe et. al., Phys. Rev. Lett. 68, 447 (1992).}
of $91\ GeV$.  A stringent constraint can also be obtained
from high-precision measurements: the result of a global
fit to all available data  concluded\Ref\PGLa{
    P. Langacker, University of Pennsylvania preprint, UPR-0492T
    (1992).}
$m_t=149^{+26}_{-31}\ GeV$.  With such a heavy mass, the
discovery of the top quark may result in a rich harvest
of physics results\rlap,\Ref\Bag{
    G. Bagliesi et. al., {\it{` Top Quark Physics-Experimental
    Aspects'}}, CERN preprint, CERN-PPE/92-05 (1992).}
and the study of top decay may even provide a window
to some new physics.
One particularly interesting example would be
 the decay $t\to H^+b$ if kinematically allowed, where
$H^+$ is a charged-Higgs-boson that occurs when
more than one (non-singlet) Higgs representation are included
in the theory.  $H^+$ must exist in a supersymmetric model,
and in many others\rlap.\Ref\kaneb{
   For a recent review see for example:  J. F. Gunion, H. E. Haber,
   G. Kane, and S. Dawson, `{\it{The Higgs Hunter's Guide}}',
   Addison-Wesley (1990).}
In this paper we wish to study
the lowest-order QCD corrections to this decay mode.

The lowest-order QCD corrections to the decay $t\to H^+b$
have been calculated before\rlap.\Ref\LiYuana{C. S. Li and T. C. Yuan,
Phys. Rev. D42, 3088 (1990).}  However,
the result of that calculation
 is erroneous.  This, for reasons to be explained below,
can be most easily seen by examining its
asymptotic behavior.  Given the potential interest of observing the
decay at high-energy colliders,
it is  necessary to have a more
careful calculation of its  order $O(\alpha_s)$
QCD corrections.
We find that for a heavy top the QCD corrections introduce
a reduction in the decay rate by about $9\%$, which
differs substantially from the earlier result of
$6\%$ of Ref. \LiYuana.

The rest of this paper is orgainized as follows.  In the
next section we discuss the
lowest-order QCD corrections to the decay $t\to bH^+$ in the
heavy top quark limit.  It is shown that up to an over all scale
 the result is the
same as for $t\to bW^+$.   The lowest-order QCD corrections to the decay
$t\to bW^+$
have been calculated independently by several groups\rlap,\REFS
\Kuhn{
  M. Jezabak and J. H. Kuhn, Nucl. Phys. B314, 1 (1989);
  {\it{ibid}}, B320, 20 (1989).}\REFSCON
  \LiuYaoa{
  J. Liu and Y. P. Yao, University of Michigan preprint,
  UM-TH-90-09 (1990) (unpublished).}\REFSCON\LiuYaob{
  J. Liu and Y. P. Yao, Int. J. Mod. Phys. A6, 4925 (1991).}
  \REFSCON\Czar{A. Czarnecki, Phys. Lett. B252, 467 (1990).}
  \REFSCON\LiYuanb{C. S. Li, R. J. Oakes, and T. C. Yuan,
  Phys. Rev. D43, 3759 (1991).}\refsend
and all agree with each other.  Thus, by employing the
equivalence theorem\Ref\ET{
J. M. Cornwall, D. N. Levin and G. Tiktopoulos,
Phys. Rev. D10, 1145 (1974);
M. S. Chanowitz and M. K. Gaillard, Nucl. Phys. B261, 379 (1985);
H. Veltman, Phys. Rev. D41, 2294 (1990);\splitout
For general gauges and subtractions see:  Y. P. Yao and C. P. Yuan,
Phys. Rev. D38, 2234 (1988).}
  the heavy-top-limit
result for $t\to bH^+$ can  be obtained from   the existing
result of Refs. \Kuhn-\LiYuanb.\
The justification for applying
the equivalence theorem to this particular calculation is also discussed.

In section 3 we provide an explicit calculation for
arbitrary $m_t$ and $m_{H^+}$, where $m_t$ and $m_{H^+}$
are the  masses of the top quark and the charged-Higgs-boson
respectively, but taking $m_b=0$ for simplicity.
The result shows the expected asymptotic behavior.
Our conclusion is given in section 4, and a few
technical details are summarized in the Appendix.
\bigskip

\chapter{Asymptotic Result}

The lowest-order QCD corrections to $t\to bH^+$
and to $t\to bW^+$ are related for the following
reasons.  Consider the heavy top quark limit.
The tree-level interaction Lagrangian for
the decay $t\to bH^+$ is given by
$$L_{H^+}=-\eta\Bigl(
{m_t\over m_{H^+}}\Bigr)\bar{b}RtH^-+h.c.,\eqno(2.1)$$
where $\eta$ is a dimensionless constant determined
by the specific theoretical model, and $R={1\over 2}(1+\gamma_5)$.
   (2.1) is a
simplification of the interactions of a large class
of theoretical models by neglecting terms directly
proportional to $m_b$.  The important feature  here
is that to the lowest-order interactions $\eta$,
$m_{H^+}$ and the Higgs field $H^+$ do not
receive QCD corrections, because to this order
the gluons only interact with the quarks.

This  feature is also shared by the
interaction Lagrangian
$$L_{\phi^+}={g\over \sqrt 2}\Bigl({m_t\over M_W}\Bigr)\bar{b}Rt\phi^-
+h.c.\eqno(2.2)$$
for the Higgs goldstone boson $\phi^+$
in the limit $m_t\gg m_b$.  As a consequence,
up to an over all scale determined by their interaction strength
difference,
the lowest-order QCD corrections to $t\to bH^+$ are
the same as to $t\to b\phi^+$ if the calculation of
the latter is carried out (to be specific) in the Feynman-'t Hooft gauge
and  $M_W$ is replaced by $m_{H^+}$.
In addition, the amplitude for
$t\to b\phi^+$  is related to that for
$t\to bW^+$
by a Ward identity, which
in the Feynman-'t Hooft gauge is
$$M_W A(t\to b\phi^+)=k^{\mu}A_{\mu}(t\to bW^+),\eqno(2.3)$$
where $A(t\to b\phi^+)$ and $\epsilon^{\mu}(k)A_{\mu}(t\to bW^+)$
are the amplitudes for $t\to b\phi^+$ and $t\to bW^+$, respectively,
and $\epsilon^{\mu}(k)$ is the $W$ polarization vector.
Thus,
knowing the QCD corrections to the Green's function $A_{\mu}$ for
the decay $t\to bW^+$ one immediately obtains the result $A$
for  $t\to b\phi^+$ from (2.3).

However, (2.3) does not  necessarily imply that
such  relations hold also
for the decay rate.
 In fact, in the absence of
CP violation the one-loop QCD corrected interaction
Lagrangian
$$L_{eff}=\bar{b}(p')\Gamma_{\mu}t(p)W^{\mu}(k)\eqno(2.4)$$
with an on-shell $b$- and $t$-quark and an arbitrary $W$
 has three independent form
factors which may be parameterized as $F_1, F_2$ and $F_3$
$$\Gamma_{\mu}={g\over \sqrt 2}\Bigl\{
F_1(k^2)\gamma_{\mu}L-N_c\Bigl({\alpha_s\over 2\pi}\Bigr)\Bigl[
F_2(k^2)i\sigma_{\mu\nu}k^{\nu}+F_3(k^2)k_{\mu}\Bigr]m_tR\Bigr\},
\eqno(2.5)$$
where $N_c=4/3$ and $\sigma_{\mu\nu}={i\over 2}[\gamma_{\mu},
\gamma_{\nu}]$.
The $F_3$ term in $A_{\mu}(t\to bW^+)$
provides a nonzero contribution to $A(t\to b\phi^+)$
via (2.3) but  not to the $t\to bW^+$ decay rate because
$\epsilon^{\mu}k_{\mu}=0$.
By contrast, the anomalous moment term
$F_2$ contributes to $t\to bW^+$ but not
to $A(t\to b\phi^+)$ because $\sigma_{\mu\nu}k^{\mu}k^{\nu}=0$.
Thus,  the complete lowest-order QCD corrections
to the decay rate for
 $t\to bW^+$ are  not the same as to that for $t\to b\phi^+$.

Nevertheless, to the leading order in $m_t$ the aforementioned difference
disappears in the limit $m_t\to \infty$.  This follows because
$m_t$ is the only heavy scale in question, and  on dimensional
grounds one has from (2.5) that
 $\lim_{m_t\to\infty}F_{2,3}/F_1=m_t^{-2}$.
Indeed, explicit calculations show
that both  $F_2$ and  $F_3$  vanish to the leading order in
$m_t$.
$F_1$ and $F_2$ have been given explicitly
in Refs. \LiuYaoa,\ \LiuYaob
$$\eqalign{&F_1(M_W^2)=1-Nc\Bigl({\alpha_s\over 2\pi}\Bigr)\Delta_0,\cr
&           F_2(M_W^2)=
-{1\over 2M_W^2}\ln\Bigl(1-{M_W^2\over m_t^2}\Bigr)
,\cr}\eqno(2.6)$$
where
$$\eqalign{\Delta_0=&2+\Bigl({3\over 2}+\ln{\mu^2\over M_W^2}\Bigr)
   \ln{m_bm_t\over m_t^2-M_W^2}+
   {1\over 2}\Bigl(\ln{\mu^2\over m_t^2}+\ln{\mu^2\over m_b^2}\Bigr)\cr
&+{1\over 2}\Bigl(\ln^2 {M_W^2\over m_t^2-M_W^2}+
 \ln^2{m_t^2\over m_t^2-M_W^2}\Bigr)
-{1\over 4}\Bigl(\ln^2{m_t^2\over M_W^2}+\ln^2{m_b^2\over M_W^2}\Bigr)
+Sp\Bigl({M_W^2\over m_t^2}\Bigr),\cr}\eqno(2.7)$$
$\mu\to 0$ is a fictitious gluon mass and
$Sp(x)=\int^1_0dy\ln y/(y-x^{-1})$ is the Spence function.
These results are valid when $m_bM_W/(m_t^2-M_W^2)\ll 1$.
$F_3$ has not been given explicitly before.   From
a straightforward calculation we find
$$F_3(M_W^2)={1\over M_W^2}\Bigl[
\Bigl({m_t^2\over M_W^2}-{3\over 2}\Bigr)\ln{m_t^2\over
m_t^2-M_W^2}-1\Bigr].\eqno(2.8)$$
{}From (2.6) and (2.8) one sees
$$\lim_{m_t\to\infty}F_2(M_W^2)=\lim_{m_t\to \infty}F_3(M_W^2)=0,
\eqno(2.9)$$
in accordance with the dimensional argument.
As a result, in this limit the rates for
$t\to bW^+$ and $t\to b\phi^+$ plus their QCD corrections
are in fact the same, and the former can be calculated from
the equivalence theorem.

The result for $t\to bW^+$ is   known (Refs. \Kuhn\  -
\LiYuanb\ )
$$\eqalign{\Gamma(t\to bW^+)=\Gamma_0(t\to bW^+)&\Bigl\{
1-N_c\Bigl({\alpha_s\over \pi}\Bigr)
\Bigl[Sp\Bigl({M_W^2\over m_t^2}\Bigr)
-Sp\Bigl(1-{M_W^2\over m_t^2}\Bigr)+{\pi^2\over 2}\Bigr]\Bigr\}\cr
-{G_Fm_t^3\over 8\sqrt 2\pi}N_c\Bigl({\alpha_s\over \pi}\Bigr)&
\Bigl\{\Bigl(1-{M_W^2\over m_t^2}-2{M_W^4\over m_t^4}\Bigr)
{M_W^2\over m_t^2}\ln{M_W^2\over m_t^2}\cr
&+{1\over 2}\Bigl(1-{M_W^2\over m_t^2}\Bigr)^2\Bigl(5+4{M_W^2\over m_t^2}
\Bigr)\ln\Bigl(1-{M_W^2\over m_t^2}\Bigr)\cr
&-{1\over 4}\Bigl(1-{M_W^2\over m_t^2}\Bigr)
\Bigl(5+9{M_W^2\over m_t^2}-6{M_W^4\over m_t^4}\Bigr)\Bigr\},\cr}
\eqno(2.10)$$
where $\Gamma_0(t\to bW^+)=G_F(m_t^2-M_W^2)^2(1+2M_W^2/m_t^2)/8\sqrt 2\pi
m_t$  is the tree-level rate.  In (2.10)  the
contribution from a virtual
gluon  exchange is
$$\Gamma(t\to bW^+)_{virt}=\Gamma_0(t\to bW^+)\Bigl\{
1-N_c\Bigl({\alpha_s\over \pi}\Bigr)\Bigl[\Delta_0-
{3\over 2}\Bigl(1+2{M_W^2\over m_t^2}\Bigr)^{-1}\ln{m_t^2
\over m_t^2-M_W^2}\Bigr]\Bigr\},\eqno(2.11)$$
where the $\Delta_0$  term is due to $F_1$, and the  last term is due to
the anomalous moment $F_2$.  The
real gluon emission  contribution to (2.10) is
$$\eqalign{\Gamma(t\to bW^+g)_{real}=
{G_FM_W^2\over 8\sqrt 2\pi m_t}\Bigl({\alpha_s\over \pi}\Bigr)
N_c&\Bigl\{{(m_t^2-M_W^2)^2\over M_W^2}\Bigl(1+2{M_W^2\over m_t^2}\Bigr)
\Bigl[\Delta_0-\ln\Bigl(1-{M_W^2\over m_t^2}\Bigr)\cr
&+Sp\Bigl(1-{M_W^2\over m_t^2}\Bigr)-Sp\Bigl({M_W^2\over m_t^2}\Bigr)
-{\pi^2\over 2}\Bigr]\cr
&+M_W^2\Bigl({m_t^2\over M_W^2}-2{M_W^2\over m_t^2}-1\Bigr)\ln
{m_t^2\over M_W^2}\cr
&+(m_t^2-M_W^2)\Bigl({27\over 12}+{5\over 4}{m_t^2\over M_W^2}
-{3\over 2}{M_W^2\over m_t^2}\Bigr)\Bigr\}.\cr}\eqno(2.12)$$
The condition for the validity of (2.12) is
$m_bm_t/(m_t^2-M_W^2)\ll 1$.

Turning back to  $t\to bH^+$, the tree-level decay rate from (2.1) is
$$\Gamma_0(t \to bH^+)={\eta^2\over 32\pi}\Bigl({m_t^2\over m_{H^+}^2}
-1\Bigr)^2\Bigl({m_{H^+}\over m_t}\Bigr)^2m_t.\eqno(2.13)$$
We already know that up to an over all scale its
lowest-order QCD correction is the same as for
$t\to b\phi^+$ (with the exchange of $M_W$ by $m_{H^+}$),
and the latter is identical to  $t\to bW^+$
in the heavy top limit.  In fact, one can show that such relations
also hold for the virtual- and real-gluon emission contributions
seperately.
{}From (2.10), we have
$$\lim_{m_t\to\infty}{\Gamma(t\to bW^+)\over
\Gamma_0(t\to bW^+)}=
1-N_c\Bigl({\alpha_s\over\pi}\Bigr)\Bigl({\pi^2\over 3}
-{5\over 4}\Bigr),\eqno(2.14)$$
where we have made use of $Sp(0)=0$ and $Sp(1)=\pi^2/6$, and hence
$$\lim_{m_t\to\infty}{\Gamma(t\to bH^+)\over
\Gamma_0(t\to bH^+)}
=1-N_c\Bigl({\alpha_s\over \pi}\Bigr)\Bigl(
{\pi^2\over 3}-{5\over 4}\Bigr).\eqno(2.15)$$
Compared with the result given by
Ref. \LiYuana\  in the limit $m_t\to \infty$,
Eq. (2.15) has a different constant,
$5/4$.  In Ref. \LiYuana\ that constant is $9/4$.

\bigskip
\

\chapter{Explicit Calculation}

In this section we  verify (2.15) by an explicit calculation.
For simplicity, we will ignore the bottom quark mass $m_b$,
but allow $m_t$ and $m_{H^+}$ to be arbitrary.

QCD corrections from a virtual gluon exchange introduce a correction to
the interaction vertex,  wave function renormalizations to
$t$ and $b$, and a mass renormalization to $m_t$.  They have
 been calculated explicitly in Ref. \LiuYaoa.\ The result is
$$\Gamma(t\to bH^+)_{virt}=
\Gamma_0(t\to bH^+)\Bigl\{
1-N_c\Bigl({\alpha_s\over\pi}\Bigr)\Bigl[
\Delta_0+\Bigl({m_t^2\over m_{H^+}^2}-{3\over 2}\Bigr)
\ln{m_t^2\over m_t^2-m_{H^+}^2}-1\Bigr]\Bigr\}.
\eqno(3.1)$$
The last
term of (2.11), which arises from the anomalous moment $F_2$,
is now replaced in (3.1) by the $F_3$ term (with
the exchange of $M_W\leftrightarrow m_{H^+}$).
As we expected, in the limit $m_t\to \infty$ both $F_2$ and $F_3$ vanish
and the QCD corrections in (2.11) and (3.1)
are the same.

The calculation for the decay $t\to bH^+g$ with a
real gluon emission is  also
straightforward.  We find (details can be found in the Appendix)
$$\eqalign{\Gamma(t\to bH^+g)_{real}=\Gamma_0(t\to bH^+)
N_c\Bigl({\alpha_s\over \pi}\Bigr)&\Bigl[\Delta_0
-\ln\Bigl(1-{m_{H^+}^2\over m_t^2}\Bigr)+Sp\Bigl(1-{m_{H^+}^2\over
\ m_t^2}\Bigr)\cr
&-Sp\Bigl({m_{H^+}^2\over m_t^2}\Bigr)
+{m_{H^+}^2\over m_t^2-m_{H^+}^2}\ln{m_t^2\over m_{H^+}^2}
+{5\over 4}-{\pi^2\over 2}\Bigr].\cr}\eqno(3.2)$$
Again,  comparing (3.2) and (2.12) we see that
in the heavy top limit their QCD corrections are indentical.
It then follows from (3.1) and (3.2) that the
final result for $t\to bH^+$ including its lowest-order
QCD corrections is
$$\eqalign{\Gamma(t\to bH^+)=\Gamma_0&(t\to bH^+)
\Bigl\{1-N_c\Bigl({\alpha_s\over\pi}\Bigr)\Bigl[
\Bigl({5\over 2}-{m_t^2\over m_{H^+}^2}\Bigr)\ln
\Bigl(1-{m_{H^+}^2\over m_t^2}\Bigr)\cr
&-{m_{H^+}^2\over m_t^2-m_{H^+}^2}
\ln{m_t^2\over m_{H^+}^2}
+Sp\Bigl({m_{H^+}^2\over m_t^2}\Bigr)-
Sp\Bigl(1-{m_{H^+}^2\over m_t^2}\Bigr)+{\pi^2\over 2}-{9\over 4}
\Bigr]\Bigr\},\cr}\eqno(3.3)$$
which is free from infrared and  collinear divergences.
It reduces to (2.15) in the limit
$m_t\to \infty$.  Although (3.3) is obtained by taking
$m_b=0$, one can show that it remains as a good approximation
as long as
$${m_bm_t\over m_t^2-m_{H^+}^2}\ll 1.\eqno(3.4)$$

Eq. (3.3) differs from the result of Ref. \LiYuana\  by a term
$-(m_t^2/m_{H^+}^2)\ln(1-m_{H^+}^2/m_t^2)$ in the
squared brackets.  This missing term approaches to $1$ in the heavy top
limit.  Numerically,
the  QCD corrections given by (3.3) turn out to be about
 $-9\%$ (for $\alpha_s=0.1$) if the top quark
is very heavy,  rather than
$-6\%$ reported in Ref. \LiYuana.
\bigskip

\chapter{Conclusion}

We have calculated the lowest-order QCD corrections to
the decay $t\to bH^+$.  A simple analytic
result is obtained for $m_bm_t/(m_t^2-m_{H^+}^2)\ll 1$.
It is shown that for a heavy top quark, the order
$O(\alpha_s)$  QCD corrections reduce the tree-level
rate of the decay $t\to bH^+$ by about $9\%$ (for $\alpha_s=0.1$)
 rather
than $6\%$ reported previously in the literature.

Following an observation that  the lowest-order
QCD corrections to the interactions $\bar{b}tH^-$  and
$\bar{b}t\phi^-$ (in the Feynman-'t Hooft gauge)  are
identical in the heavy top limit up to an over all scale, we have shown
that asymptotically the lowest-order QCD corrections
to $t\to bH^+$ and to $t\to bW^+$ are the same again
up to an over all scale.
We also verified explicitly that the anomalous form factors $F_2$ and
$F_3$ vanish in  leading order, and as a result
 the leading term of the
QCD corrections to $t\to bW^+$ in the heavy top limit
can  be
calculated from the equivalence theorem.
\bigskip

\ack
We wish to thank Paul Langacker for valuable discussions and comments.
This work was supported in part by the U. S. Department
of Energy, contract DE-AC02-76-ERO-3071 (J. L.) and
DE-AC02-76-ERO-1112 (Y. P. Y.).
\endpage
\centerline{APPENDIX}

In this Appendix we give some details for the calculation
of $\Gamma(t\to bH^+g)_{real}$.  The
result for this decay with only a soft gluon emission
is known (Ref. \LiuYaoa).  The rate
for $t\to bH^+g$ with a hard gluon emission has also
been calculated numerically before\rlap.\Ref\Samuel{
  J. Reid, G. Tupper, G. Li, and M. S. Samuel, Z. Phys. C51, 395 (1991).}
These result are  sensitive to the infrared- and the
collinear-cut determined by the experiment apparatus.
Here we will present an analytic calculation that
 takes both the soft and the hard gluon emission
into account.  The calculation will be carried out in
the limit $m_b=0$.  The condition for the validity of its
result can be extended to that given by (3.4) of the text.

The matrix element of the decay $t(p)\to b(p')H^+(k)g(q)$ is
$$M(t\to bH^+g)=\eta g_s\epsilon^{\nu}(q)\Bigl({m_t\over m_{H^+}}\Bigr)
\bar{u}_b(p')\Bigl[R{1\over p\!\!\!/-q\!\!\!/-m_t}\gamma_{\nu}
+\gamma_{\nu}{1\over p\!\!\!/'+q\!\!\!/-m_b}R\Bigr]u_t(p),
\eqno(A.1)$$
where for simplicity we have not displayed the color matrix
$\lambda/2$ explicitly.  The spin-summed matrix square is
$$-{N_c\eta^2g^2\over 8m_tm_b}\Bigl({m_t\over m_{H^+}}\Bigr)^2
\Bigl[I_1+I_2+I_3\Bigr],\eqno(A.2)$$
where
$$\eqalign{I_1={1\over [(p-q)^2-m_t^2]^2}\Bigl[&
-8[p'\cdot (p-q)][p\cdot (p-q)]+4(p-q)^2(p\cdot p')\cr
&+16m_t^2[p'\cdot (p-q)]-4m_t^2(p\cdot p')\Bigr],\cr
I_2={1\over [(p'+q)^2-m_b^2]^2}\Bigl[&
-8[p'\cdot (p'+q)][p\cdot (p'+q)]+4(p'+q)^2(p\cdot p')\cr
&+16m_b^2[p\cdot (p'+q)] -4m_b^2(p\cdot p')\Bigr],\cr
I_3={1\over[(p-q)^2-m_t^2][(p'+q)^2-m_b^2]}&\Bigl[
16[p\cdot (p'+q)][p'\cdot (p-q)]-8m_t^2[p'\cdot (p'+q)]\Bigr].\cr}
\eqno(A.3)$$
It then follows that
$$\Gamma(t\to bH^+g)_{real}=-{N_c\eta^2g_s^2\over 4(2\pi)^5m_t}
\Bigl({m_t\over m_{H^+}}\Bigr)^2\Bigl[\Gamma_1+
\Gamma_2+\Gamma_3\Bigr],\eqno(A.4)$$
where
$$\Gamma_{1,2,3}=\int{d^3\vec{p}'\over 2p'_0}
                     {d^3\vec{q}\over 2q_0}
                     {d^3\vec{k}\over 2k_0}\delta^4(p-p'-q-k)
                     I_{1,2,3}.\eqno(A.5)$$

To evaluate (A.5), we employ the standard method of decomposing
a three-body phase space integral into products of
two-body phase space integrals.  Introducing a fictitious gluon mass
$\mu$ to regularize the infrared singularity , we find
$$\eqalign{\Gamma_1
={\pi^2\over 4m_t^2}&\int^{(m_t-\mu)^2}_{(m_{H^+}+m_b)^2}
da^2{\lambda^{1/2}(m_t^2,a^2,\mu^2)\lambda^{1/2}(a^2,m_{H^+}^2,
m_b^2)\over a^2(a^2-m_t^2)^2}\cr
&\ \ \ \ \ \times\Bigl[
-2(a^2-m_{H^+}^2)(a^2+m_t^2)+8m_t^2(a^2-m_{H^+}^2)
+{(a^2-m_{H^+}^2)(a^4-m_t^4)\over a^2}\Bigr]\cr
=-{\pi^2\over m_t^2}&\Bigl[(m_t^2-m_{H^+}^2)^2\Bigl(
1+\ln{\mu m_t\over m_t^2-m_{H^+}^2}\Bigr)-{m_t^2m_{H^+}^2\over 2}
\Bigl(1+{5\over 2}{m_{H^+}^2\over m_t^2}\Bigr)\ln{m_t^2\over m_{H^+}^2}\cr
&\ \ \ +{m_t^2-m_{H^+}^2\over 4}\Bigl({5\over 2}m_{H^+}^2
+{9\over 2}m_t^2\Bigr)\Bigr],\cr}\eqno(A.6)$$
where  $\lambda(x,y,z)=x^2+y^2+z^2-2(xy+xz+yz)$.
Also,
$$\eqalign{\Gamma_2
={\pi^2\over 4m_t^2}&\int^{(m_t-m_{H^+})^2}_{(m_b+\mu)^2}db^2
{\lambda^{1/2}(m_t^2,b^2,m_{H^+}^2)\lambda^{1/2}(b^2,m_b^2,\mu^2)\over
b^2(b^2-m_b^2)^2}\cr
&\ \ \ \ \ \times(m_t^2+b^2-m_{H^+}^2)\Bigl[
-2b^2+6m_b^2+{b^4-m_b^4\over b^2}\Bigr]\cr
=-{\pi^2\over m_t^2}&\Bigl[(m_t^2-m_{H^+}^2)^2\Bigl(
1+{1\over 2}\ln{\mu^2(m_t^2-m_{H^+}^2)\over m_tm_b^3}\Bigr)\cr
&-{1\over 4}(2m_t^2m_{H^+}^2-m_{H^+}^4)\ln{m_t^2\over m_{H^+}^2}
-{3\over 8}(m_t^2-m_{H^+}^2)\Bigl(m_t^2-{5\over 3}m_{H^+}^2\Bigr)
\Bigr].\cr}\eqno(A.7)$$
The calculation of $\Gamma_3$ is most complicated, we find
$$\eqalign{\Gamma_3={\pi^2\over m_t^2}&
\int^{(m_t-m_{H^+})^2}_{(m_b+\mu)^2}db^2
{\lambda^{1/2}(m_t^2, b^2, m_{H^+}^2)\over b^2-m_b^2}\Bigl\{
{\lambda^{1/2}(b^2,m_b^2,\mu^2)\over b^2}(m_t^2-m_{H^+}^2+b^2)\cr
&+\Bigl[(m_t^2-m_{H^+}^2+b^2)(m_t^2-m_{H^+}^2+m_b^2)-m_t^2
(b^2+m_b^2)\Bigr]J_0\Bigr\},\cr}\eqno(A.8)$$
where
$$\eqalign{&J_0=
{1\over\lambda^{1/2}(m_t^2,b^2,m_{H^+}^2)}\cr
&\times\ln{(b^2-m_b^2+\mu^2)(m_t^2+b^2-m_{H^+}^2)-\lambda^{1/2}
(m_t^2,b^2,m_{H^+}^2)\lambda^{1/2}(b^2,\mu^2,m_b^2)\over
(b^2-m_b^2+\mu^2)(m_t^2+b^2-m_{H^+}^2)+\lambda^{1/2}
(m_t^2,b^2,m_{H^+}^2)\lambda^{1/2}(b^2,\mu^2,m_b^2)}.\cr}\eqno(A.9)$$
Neglecting terms of the order of and smaller than
$m_bm_t/(m_t^2-m_{H^+}^2)$, we find from (A.8) and (A.9)
$$\eqalign{\Gamma_3=-{\pi^2\over m_t^2}&
\Bigl\{(m_t^2-m_{H^+}^2)^2\Bigl[
\Bigl(2+\ln{\mu^2\over m_{H^+}^2}\Bigr)\ln
{m_bm_t\over m_t^2-m_{H^+}^2}-{1\over 4}
\Bigl(\ln^2{m_b^2\over m_{H^+}^2}+\ln^2{m_t^2\over m_{H^+}^2}\Bigr)\cr
&+{1\over 2}\Bigl(\ln^2{m_t^2\over m_t^2-m_{H^+}^2}
                 +\ln^2{m_{H^+}^2\over m_t^2-m_{H^+}^2}\Bigr)
+Sp\Bigl(1-{m_{H^+}^2\over m_t^2}\Bigr)-{\pi^2\over 2}+1\Bigr]\cr
&+2m_t^2m_{H^+}^2\ln{m_t^2\over m_{H^+}^2}
-{1\over 2}m_t^4\Bigl(1-{m_{H^+}^2\over m_t^2}\Bigr)
\Bigl(1+3{m_{H^+}^2\over m_t^2}\Bigr)\Bigr\}.\cr}\eqno(A.10)$$
Substituting (A.6), (A.7) and (A.10) into (A.4), we obtain
the result given by (3.2) of the text.
\endpage
\refout
\end